# Light collection of POLAR detector

Dong YongWei[†], Li YanGuo, Wu BoBing, Zhang YongJie, Zhang Shuangnan

Key Laboratory of Particle Astrophysics, Institute of High Energy Physics, Chinese Academy of Sciences, Beijing 100049, China

**POLAR is a compact polarimeter dedicated to measure the polarization of GRBs between 50-300 keV. The light collection of 200\*6\*6mm³ plastic bars has been simulated and optimized in order to get uniform response to x-rays at different points of one single bar. According to the Monte Carlo results, the amplitude uniformity strongly depends on the polishing level of scintillator surface and the covering. A uniformity of 89% is achieved with a prototype constructed by a non position-sensitive PMT and an array of 4×4 bars.**

Plastic scintillator, light collection, Gamma-ray Burst, Polarimeter

Received May 20, 2009; accepted
doi:
[†]Corresponding author (email:) dongyw@ihep.ac.cn
Supported by National Basic Research Program of China - 973 Program 2009CB824800

Gamma-Ray Bursts (GRB), the brightest explosions in the universe, remains one of the most interesting topics in high energy astrophysics. While the spectral and timing analysis of GRBs allow several different models to explain the observation data. The origin of energetic gamma-ray bursts is still unknown. The detection of polarization of gamma-ray may provide insight into the mechanism driving these powerful explosions since the polarization levels predicted by different GRB models are clearly different. The synchrotron with ordered magnetic field model can produce a linear polarization of 20%~70%, while the synchrotron with small-scale random magnetic field model and the Compton drag model will lead to < 20% typically. The maximum polarization can reach 100% for the Compton drag model while 70% for the synchrotron with small-scale random magnetic field model [1]. POLAR was then proposed as a dedicated Compton polarimeter for GRBs [2]. It is scheduled to be launched onboard the Chinese Space Laboratory in 2012.

The POLAR detector is made up of 4×5 identical detector units each consisting of 8×8 organic scintillator bars. One detector unit is coupled to an H8500 MAPMT and readout by a VA32/TA32 ASIC chip. One piece of optical grease is inserted between the plastic bars and the PMT acting as shock absorption layer. The front and lateral sides of the array are covered by a 2 mm thick Al layer to absorb the low energy photons and charged particles. The schematic drawing of one POLAR detector unit is shown below (see Fig. 1). Each plastic bar (BC-448) is 6 mm×6 mm×200 mm matching the pixel of H8500.

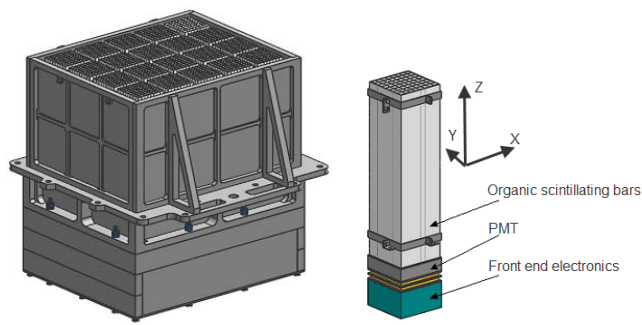

**Fig. 1 Left: mechanical design of POLAR payload. Right: A detector unit of POLAR.**

According to the principle of Compton polarimetry[3], The differential Compton cross-section of polarized photon is angular dependent, which is expressed as [4]:

$$\frac{d\sigma}{d\Omega} = \frac{r_0^2 \varepsilon^2}{2}(\frac{1}{\varepsilon} + \varepsilon - 2\sin^2\theta \cos^2\eta),$$

where $r_0$ is the classical electron radius, $\varepsilon$ is the ratio of recoil energy to incident energy, $\theta$ is the angle between the incident photon direction and the scattered photon direction and $\eta$ is the azimuthal angle of the scattered photon with respect to the electric vector of the incident photon. Asymmetry in the distribution of recoil photons can be used to determine the direction and degree of the polarization of GRBs.

For each Compton event in POLAR detector, the projection of scattered photon on the detector plane can be detected by the bars with the two highest energy depositions [2]. An effective event is defined when at least two coincident interactions with an energy deposition larger than the trigger threshold. The two highest energy depositions should occur in non-adjacent bars and the total energy deposition should be less than 500 keV. The localization of GRBs can be derived from other missions through the GCN (The GRB Coordinates Network). Based on the simulation, about 10 GRBs/yr can be detected with an MDP (Minimum Detectable Polarization) < 10% [5]. Thus POLAR is sufficient to distinguish between the theoretical models for the prompt emission of GRBs.

## 1 Light collection and uniformity of amplitude

The same energy deposition in different points of one bar may not result in the same amplitude of the output signals. Bad uniformity for one single bar could reduce detection efficiency. A beam of 60 keV photons illuminated from the top of POLAR is taken as an example. The Compton edge of a 60keV photon in BC448 is ~11keV. If the trigger threshold is 5 keV, a 20% drop in amplitude will lose all the Compton events with energy deposition between 5 keV and 6.25 keV. So 21% of the effective events are lost due to bad uniformity of amplitude. No fluctuation is considered. Great efforts shall be taken to increase the uniformity of amplitude.

The POLAR plastic bar is long and narrow, which greately deteriorate the efficiency of light collection. Such bars are usually used in Time-Of-Flight detectors. While the long plastic bar in TOF is to get good time resolution, not to get uniform amplitude [6]. So the optical design of polar would be slightly different.

The incident X-rays from GRBs deposit all or part of its energy in the plastic bar. Numerous scintillation photons are generated, the number of which is determined by the light yield (~11 photons per keV). For this process the fluctuation obeys the Fano process with the FANO factor of ~0.14. The next processes, which were named together as transfer and collection processes by Breitenberger [7], include incomplete collection of the emitted photons at the PMT photocathode, production of photoelectrons in response and the collection of these photoelectrons at the first dynode of the PMT. The final process is electron multiplication. The processes of photoelectron collection and the multiplication are ignored in this paper.

There are several endings of the scintillation photons:
  (1) Absorbed in the medium;
  (2) Absorbed at the boundary between the medium and the covering (or PMT cathode);
  (3) Detected by the PMT cathode;
  (4) Escaping from the light transfering system.

The probability of (1) depends on the light attenuation length of the mediums. For a plastic scintillator of BC408 series, the bulk attenuation length is around 250cm. the ratio of the first ending would be very small. The Q.E. curve of PMT defines whether the photon is DETECTED or ABSORBED. A tiny part of the photons will escape from the system because the PMT cathode doesn't match quite well with the plastic bar in dimension. It will be one cause of crosstalk between pixels.

The uniformity of amplitude can be measured in two ways:

(1) To compare the full-energy peak position (or mid-position of Compton edge) with x-ray illuminated at different positions of the plastic bar.
(2) To compare the ratio of collected photoelectrons to optical photons generated by the plastic scintillator.

Obviously, it is more accurate using the 2$^{nd}$ method without fluctuation of the number of scintillation photons, though the 2$^{nd}$ method can't be realized by experiments. The ratio indicated in 2nd method is named as the Light Collection Efficiency (LCE) hereafter.

## 2 Monte Carlo simulation

Several Monte Carlo program may be used for the light collection simulation, e.g. GEANT4 [8], DETECT2000 [9], PHOTRACK [10]. Here we use the optical package in GEANT4.

A UNIFIED model is defined for a rough surface of scintillator. It allows four constants to control the radiant intensity of the surface: $C_{sl}$(the specular lobe constant), $C_{ss}$(the specular spike constant), $C_{bs}$(the backscattering constant) and $C_{dl}$(the diffuse lobe constant). The four constants are added up to 1. Actually $C_{ss}$ represents the fraction of specular reflection and $C_{dl}$ represents the fraction of diffuse reflection. $C_{sl}$ and $C_{bs}$ represent the degree of roughness in other ways. So $C_{ss}$ is 1 for a theoretical polished surface and $C_{dl}$ is 1 for a 100% percent rough surface.

The observed field angle of PMT cathode is very small at the far end of the bar. Less than 1% of the total photons can directly reach the PMT without reflections. There should be a number of reflections along the travel of most detected photons. So it is quite reasonable to choose a high-reflectivity covering. A diffuse reflector ( "backpainted" model in G4), a specular reflector or a diffuse paint ("frontpainted" model) may be the optional covering of the plastic bar. They all show good ability in reflecting the photons back into the original medium. With the parameters of plastic bar and the covering defined, the light transfering system is then established.

General response to x-ray is firstly simulated. About 10% of the 60 keV x-ray lose their energy in the bar, see fig. 2. The counts in full-energy peak is 3% of the overall incident x-rays. The energy resolution is 32% @60keV with only fluctuation of scintillation process and LCE process considered.

Using the 2$^{nd}$ method mentioned in the last section, an average LCE is derived. The photoelectron yield of 1keV deposited energy is ~1p.e and the LCE is ~9% in this condition. Simply taking the Q.E. of PMT as 25%, there are altogether 36% of scintillation photons finally arriving at the cathode surface. The trigger threshold of POLAR shall not be smaller than 5 keV to distinguish effective events from the dark noise of PMT.

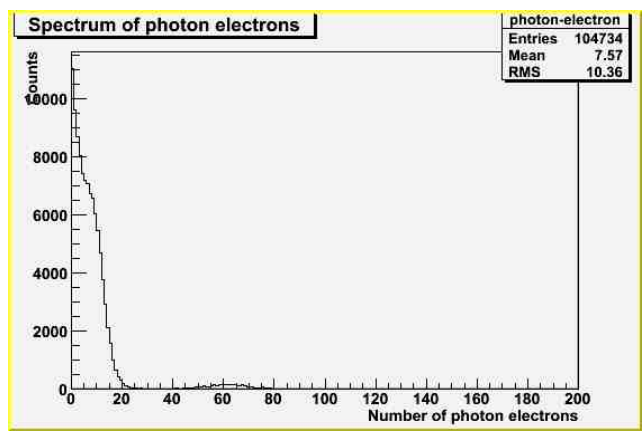

**Figure 2** the photoelectron spectrum of BC448. 1e6 photons of 60 keV are emitted at the nearside of the bar in +X direction. The plastic bar is covered by a specular reflector (Reflector Coefficient = 0.98). The FANO factor is set to 1.

The solutions of different reflectors are compared in fig. 3. The TEFLON material is a covering of diffuse reflection (RC = 0.95). Though the solution of polished surface and diffuse reflectors achieves the highest LCE at the point 6mm away from the PMT, LCE drops quickly with the distance. The best uniformity of TEFLON solution is 81.5% which is clearly unacceptable.

ESR is an ideal material of specular reflection. It hasreflectivity as high as 0.985 and no thermal distortion even in liquid nitrogen temperature. The combination of polished surface and ESR covering has the best uniformity of amplitude than other solutions. An easy explanation is that a small addition of diffuse reflection would increase the probability of photons remaining at the original region. Rough surface or diffuse reflector used at the far end of the bar is the reason preventing photons from reaching the PMT.

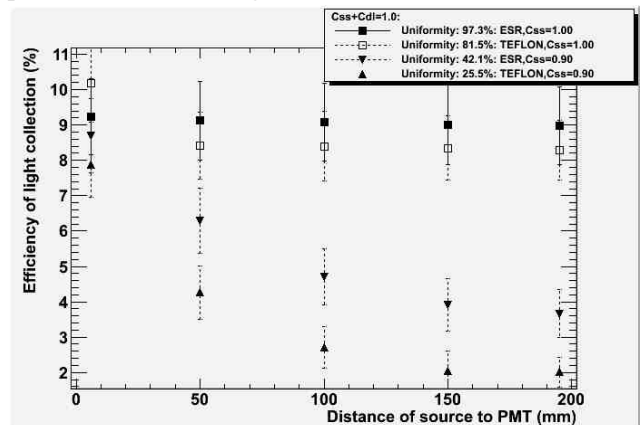

**Figure 3** Dependence between LCE and the distance of source to PMT. Different coverings and different polishing level for the scintillator surface are compared.

The polishing level of the scintillator surface is another key parameter affecting the LCE. $C_{ss}$ represents the degree of polish as indicated in fig. 4. The drop rate of LCE at the far end of the bar is much faster than at the nearside no matter what kind of covering is used. Thus a common conclusion can be drawn that the uniformity drops quickly with the decreasing of specular reflection coefficient $C_{ss}$. $C_{sl}$ and $C_{bs}$ have similar effects on the LCE.

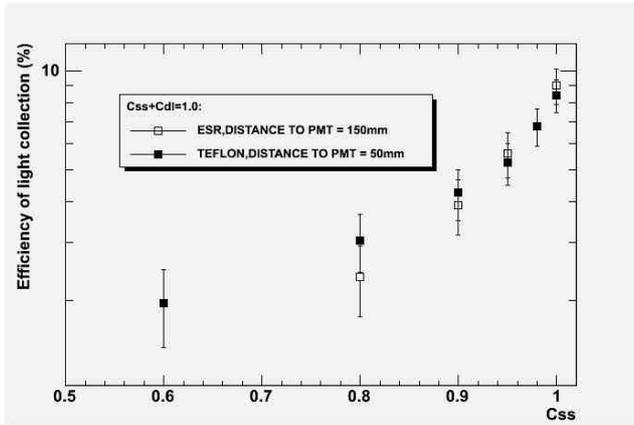

**Figure 4** Dependence between LCE and polishing level of scintillator surface.

We are surprised to see that the uniformity doesn't vary much with the RC value for a 100% polished surface. With RC changes from 0.985 to 0.85, the uniformity drops from 97.0% to 94.6%. An alternative material of specular reflection like aluminum foil may also achieve good uniformity. A side effect is that the average amplitude becomes smaller than ESR solution.

## 4 Experiment results

According to the simulation results, an optimized design of POLAR detector is deduced. A prototype with 4×4 plastic bars has been established based on polished bars and ESR covering.

The machining procedure of the bars shall be taken good care of to avoid significant heat or mechanical stress. One sheet of flat scintillator is cut into separate bars with the minimum tolerence. Those bars having visible defects and large tolerences are not selected. The polishing procedure makes use of the ultra fine sandpaper. We can't make 100% polished surface in practice, especially for plastic scintillators.

Two identical ESR sheets, each properly cut along the height in order to interlock with each other, form a grid to install the 16 bars. A PMT of $\varphi$=50mm is coupled to the detection unit. A radiation experiment was performed to test the uniformity after integration and encapsulation.

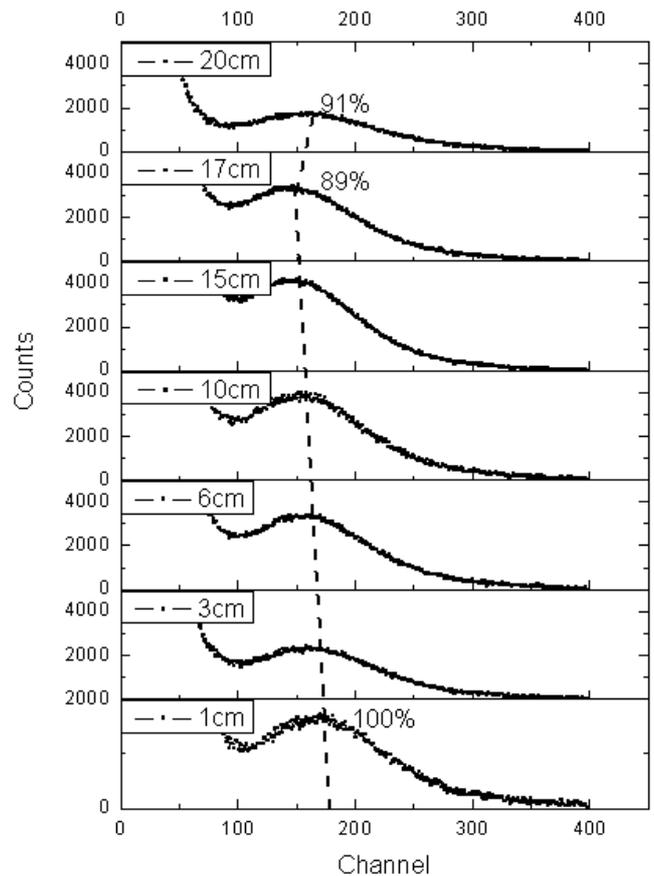

**Figure 5** Energy spectrum of the detection unit. The length in cm represent distance of sources to PMT. a $^{241}$Am radioactive source is placed at different points on the lateral side pointing the +X direction.

The 1$^{st}$ method in section 1 is used because the full-energy peaks can be well fit by Gauss function. The spectrums result in an amplitude uniformity of 89%. Only 9% of effective events are lost according to the estimation method in the first paragraph of section 1.

## 5 Conclusions

Light collection in long plastic bars is discueed to promote POLAR performance. We have compared the amplitude uniformity for solutions with different scintillator surface and different covering. The best uniformity is given by polished surface and ESR reflector. An amplitude uniformity of 89% is realized in lab and acceptable in detection efficiency. More work on the trigger strategy and the crosstalk is being carried out.